\documentclass[a4paper,12pt]{article}
\usepackage{graphicx}
\usepackage{color}
\begin{document}

%--#[ Title page :
\begin{titlepage}
\vspace*{0.1cm} \rightline{KU-PH-013}
\vspace*{0.1cm}\rightline{TTP12-042}
\vspace*{0.1cm}\rightline{SFB/CPP-12-85}

\vspace{1mm}
\begin{center}

{\Large{\bf Full $\mathcal{O}(\alpha)$ electroweak radiative corrections to                                                                                  
$e^+e^- \rightarrow t \bar{t} \gamma$ with GRACE-Loop  }}

\vspace{1.5cm}
P.H. Khiem$^{A,B}$, J. Fujimoto$^{A}$, T. Ishikawa$^{A}$, T. Kaneko$^{A}$,\\
 K. Kato$^C$, Y. Kurihara$^{A}$, \\
Y. Shimizu$^A$, T. Ueda$^{D}$, J.A.M. Vermaseren$^E$, Y. Yasui$^{F}$ \\
\vspace{4.5mm}

\textit{ $^{A)}$KEK, Oho 1-1, Tsukuba, Ibaraki 305-0801, Japan.\\ 
  $^{B)}$SOKENDAI University, Shonan Village, Hayama, Kanagawa 240-0193 Japan. \\ 
  $^{C)}$Kogakuin University, Shinjuku, Tokyo 163-8677, Japan.\\ 
  $^{D)}$Karlsruhe Institute of Technology (KIT), D-76128 Karlsruhe, Germany. \\
  $^{E)}$NIKHEF, Science Park 105, 1098 XG Amsterdam, The Netherlands.\\ 
  $^{F)}$Tokyo Management College, Ichikawa, Chiba 272-0001, Japan. }\\

\vspace{10mm}
\abstract{We present the full $\mathcal{O}(\alpha)$ electroweak radiative 
corrections to the process $e^+e^- \rightarrow t \bar{t} \gamma$ at the 
International Linear Collider (ILC). The computation is performed with the 
help of the GRACE-Loop system. We present the total cross-section and the 
top quark forward-backward asymmetry ($A_{FB}$) as a function of the 
center-of-mass energy and compare them with the process $e^+e^- \rightarrow 
t \bar{t}$. We find that the value of 
$A_{FB}$ in $t \bar{t} \gamma$ production is larger than $A_{FB}$ in 
$t\bar{t}$ production. It is an important result for the measurement of the 
top quark forward-backward asymmetry at the ILC. Applying a structure 
function method, we also subtract the QED correction to gain the 
genuine weak correction in both the $\alpha$ scheme and the $G_{\mu}$ 
scheme ($\delta_{W}^{G_{\mu}}$). We obtain numerical values for 
$\delta_{W}^{G_{\mu}}$ which are changing from $2\%$ to $-24\%$ when 
we vary the center-of-mass energy from $360$ GeV to $1$ TeV.}

\end{center}
\end{titlepage}

%--#] Title page :
%--#[ Introduction :

\section{Introduction}

The experimental results of CDF~\cite{Aaltonen:2011kc} and 
D$0$~\cite{Abazov:2007ab} on the measurement of top pair production at the 
Tevatron show an unexpected large top quark forward-backward asymmetry. 
The precise theoretical calculations of the top pair production play an 
important role in explaining the experimental data. QCD radiative 
corrections to top pair production from proton-proton collisions were 
calculated by several authors~\cite{Denner:2010jp},~\cite{Denner:2012yc},~\cite{Bevilacqua:2010qb},
\cite{Melnikov:2011ta},~\cite{Melnikov:2011qx}.
 However, the measurement is affected by a 
huge background from QCD. A good example is the $gg\rightarrow t\bar{t}$ 
reaction.  In the future, the measurement will be performed at the ILC without QCD background.
 Therefore, we consider the precise calculations of 
top pair production and top pair with photon production in $e^+ e^-$ 
collisions. A completed full one-loop electroweak correction calculation to 
the process $e^+e^- \rightarrow t \bar{t}$ has already been presented in 
refs~\cite{Fujimoto:1987hu},~\cite{Fleischer:2002rn},~\cite{Fleischer:2003kk}. 
In this paper, we calculate the full $\mathcal{O}(\alpha)$ electroweak 
radiative corrections to both the process $e^+e^- \rightarrow t \bar{t}$ 
and $e^+e^- \rightarrow t \bar{t} \gamma$ at the ILC.

The data of the ATLAS~\cite{:2012gk} and CMS~\cite{:2012gu}
experiments prove the existence of a new boson with mass around $126$GeV. 
It is assumed to be the standard-model Higgs particle. Once the discovery 
of the Higgs boson is confirmed, the next important task is to measure its 
properties. However, it is clear that such a measurement is much easier at 
the cleaner environment of the ILC than at the LHC with its large QCD 
backgrounds. To measure the properties of the new boson it is important 
that also the radiative correction calculations for the ILC take the complete 
standard model into account.

The experiments at the ILC require a precise determination of the 
luminosity which will be based on higher order theoretical calculations of 
the Bhabha scattering cross-section. Thus, the computation of electroweak 
radiative corrections to the process $e^+e^- \rightarrow e^+e^- \gamma$ is 
mandatory. This will be our eventual target. However, as a first step, we 
are going to calculate the process $e^+e^- \rightarrow t \bar{t} \gamma$ 
which is easier in several respects: there are fewer diagrams and the 
numerical cancellations between the diagrams are less severe. It will 
provide a framework for our target calculation.

In the scope of this paper, we discuss the full $\mathcal{O}(\alpha)$ 
electroweak radiative corrections to the process $e^+e^- \rightarrow t 
\bar{t} \gamma$ at ILC. We then examine the numerical results of the top 
quark forward-backward asymmetry as well as the genuine weak corrections in 
both the $\alpha$ scheme and the $G_{\mu}$ scheme \cite{Fleischer:1988kj} as compared to the 
process $e^+e^- \rightarrow t \bar{t}$.

The paper is organized as follows. In section 2 we introduce the 
GRACE-Loop system and set up the calculation. In section 3 we discuss 
the numerical results of the calculation.
%  of the total cross section and top quark forward-backward asymmetry as 
%  a function of the center-of-mass energy.
%  We also explain how to extract the genuine electroweak corrections from
%  the full corrections by using the QED structure function method in this
%  section. 
Future plans and conclusions of our paper are presented in section 4.

%--#] Introduction :
%--#[ GRACE-Loop :
  
\section{GRACE-Loop and the process $e^- e^+\rightarrow t\bar{t}\gamma$}

\subsection{GRACE-Loop}

The computation is performed with the help of the GRACE-Loop system which 
is a generic program for the automatic calculation of scattering processes 
in High Energy Physics. Of course, with a system this complicated still 
under development, it is important to have as many tests as possible of the 
correctness of the answer. Hence the GRACE-Loop system has been equipped 
with non-linear gauge fixing terms in the Lagrangian which will be described 
in some of the next paragraphs. The renormalization has been carried out with
the on-shell renormalization condition of the Kyoto scheme, as described in 
ref~\cite{kyotorc}. The program was presented and checked carefully with a 
variety of $2\rightarrow 2 $-body electroweak processes in 
ref~\cite{Belanger:2003sd}. The GRACE-Loop system has also been used to 
calculate $2\rightarrow 3$-body processes such as $e^+e^- \rightarrow 
ZHH$~\cite{Belanger:2003ya}, $e^+e^- \rightarrow t \bar{t} 
H$~\cite{Belanger:2003nm}, $e^+e^- \rightarrow \nu \bar{\nu} 
H$~\cite{Belanger:2002me}. The above calculations have been done independently by
other groups, for example the process $e^+e^- \rightarrow 
ZHH$~\cite{Zhang:2003jy}, $e^+e^- \rightarrow t \bar{t} H$~\cite{You:2003zq},~\cite{Denner:2003ri},~\cite{Denner:2003zp}
and $e^+e^- \rightarrow \nu \bar{\nu} H$~\cite{Denner:2003yg},~\cite{Denner:2003iy}. Moreover, the $2\rightarrow 4$-body process as
$e^+e^- \rightarrow \nu_{\mu}  \bar{\nu}_{\mu} HH$~\cite{Kato:2005iw} was calculated successfully by GRACE-loop system.
 
The steps of calculating a process in the GRACE system are as follows. 
First the system requires input files that describe the Feynman rules of 
the model. In this case, we use the standard model. These files 
are considered part of the system but for different models the user would 
have to provide them. Next a (small) file is needed that selects the model, 
the names of the incoming and outgoing particles, and one of a set of 
predefined kinematic configurations for the phase space integration. In the 
intermediate stage symbolic manipulation handles all Dirac and tensor 
algebra in $n$-dimensions, reduces the formulas to coefficients of tensor 
one-loop integrals and writes the formulas in terms of FORTRAN subroutines 
on a diagram by diagram basis. For this manipulation either FORM~\cite{form} 
or REDUCE~\cite{reduce} is used. The FORTRAN routines will be combined with 
libraries which contain the routines that reduce the tensor one-loop 
integrals into scalar one-loop integrals. The scalar one-loop integrals 
will be numerically evaluated by one of the FF~\cite{ff} or 
LoopTools~\cite{looptools} packages. The ultraviolet divergences 
(UV-divergences) are regulated by dimensional regularization and the 
infrared divergences (IR-divergences) will be regulated by giving the 
photon an infinitesimal mass $\lambda$. Eventually all FORTRAN 
routines are compiled and linked with the GRACE libraries which include the 
kinematic libraries and the Monte Carlo integration program 
BASES~\cite{bases}. The resulting executable program can then calculate 
cross-sections and generate events.

Ref~\cite{Belanger:2003sd} describes the method used by the GRACE-Loop 
system to reduce the tensor one-loop five- and six-point functions into 
one-loop four-point functions.

The GRACE-Loop system allows the use of non-linear gauge fixing 
conditions~\cite{nlg-generalised} which are defined by:
\begin{eqnarray}
  {{\cal L}}_{GF}&=&-\frac{1}{\xi_W}
|(\partial_\mu\;-\;i e \tilde{\alpha} A_\mu\;-\;ig c_W
\tilde{\beta} Z_\mu) W^{\mu +} + \xi_W \frac{g}{2}(v
+\tilde{\delta} H +i \tilde{\kappa} \chi_3)\chi^{+}|^{2} \nonumber \\
& &\;-\frac{1}{2 \xi_Z} (\partial\cdot Z + \xi_Z \frac{g}{ 2 c_W}
(v+\tilde\varepsilon H) \chi_3)^2 \;-\frac{1}{2 \xi_A} (\partial \cdot A
)^2 \;.
\end{eqnarray}
% We are working in the 't Hooft-Feynman gauge by choosing 
% $\xi_{W}=\xi_{Z}=\xi_{A}=1$. 
We are working in the $R_{\xi}$-type gauges with condition $\xi_{W}=\xi_{Z}=\xi_{A}=1$ 
(with so-called  the 't Hooft-Feynman gauge),
 there is no contribution of the 
longitudinal term in the gauge propagator. This choice has not only the 
advantage of making the expressions much simpler. It also avoids  
unnecessary large cancellations, high tensor ranks in the one-loop 
integrals and extra powers of momenta in the denominators which cannot be 
handled by the FF package.

The GRACE-Loop system can also use an axial gauge for external photons. 
This has two advantages.
\begin{enumerate}
\item It cures a problem with large numerical cancellations. 
This is very useful when calculating the process at small angle and 
 energy cuts of the final state particles.
\item It provides a useful tool to check the consistency of the results 
which, due to the Ward identities, must be independent of the choice of the 
gauge.
\end{enumerate}

%--#] GRACE-Loop :
%--#[ ee->ttg :

\subsection{The numerical test of the process $e^- e^+\rightarrow t\bar{t}\gamma$}

\begin{figure}[h]
% \begin{center}
\hspace*{-0.5cm}\includegraphics[width=6.2in, height=12.05cm]{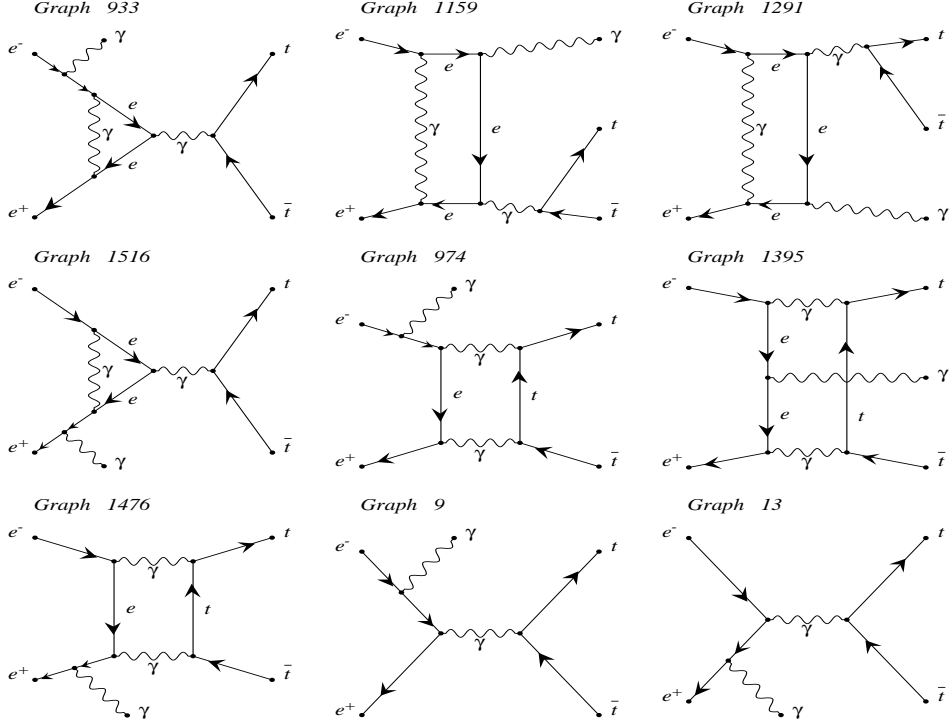}
% \end{center}
\caption{\label{feynmandiagrams} Typical Feynman diagrams as 
generated by the GRACE-Loop system.}
\end{figure}

The full set of Feynman diagrams with the non-linear gauge fixing as 
described before consists of $16$ tree diagrams and  $1704$ one-loop 
diagrams (of which 168 are pentagon diagrams). In Fig~\ref{feynmandiagrams} 
we show some selected diagrams.

The results are checked carefully by three kinds of consistency tests. 
These tests are performed with quadruple precision at a few random points in 
the phase space. The first test is ultraviolet finiteness of the 
results. This test is done on all virtual one-loop diagrams and their 
counter terms and we treat 
$C_{UV}=1/\epsilon -\gamma_{E}+\log4\pi$ as a parameter. In order to 
regularize the infrared divergences, we give the virtual photon a 
fictitious mass, $\lambda=10^{-17}$GeV. In the table~\ref{cuv} we 
present the numerical results of the test at one random point in the phase space.
 The result is stable over more than 30 digits for various values of the ultraviolet parameter.

\begin{table}[h]
\centering
\begin{tabular}{cc} \hline 
$C_{UV}$ & $2\mathcal{\Re}(\mathcal{T}_{Tree}^{+}\mathcal{T}_{Loop})$\\ \hline  \hline 
$0$     & $-5.3131630854021768119477116628317605E^{-3} $\\ \hline 
$10$    & $-5.3131630854021768119477116628317726E^{-3} $\\ \hline 
$100$   & $-5.3131630854021768119477116628320404E^{-3} $\\ \hline 
\end{tabular}
\caption{\label{cuv} Test of $C_{UV}$ independence of the amplitude. 
In this table, we take the non-linear gauge parameters to be $0$, 
$\lambda=10^{-17}$GeV and we use $1$ TeV for the center-of-mass energy.}
\end{table}

The second test is the independence of the result on the fictitious photon 
mass $\lambda$. In this case, we take $C_{UV} = 0$.  This test will be 
performed by including as well the virtual loop diagrams as the soft 
bremsstrahlung contribution. In table~\ref{lambda} the numerical 
results of the test are presented. We find that the result is stable over 
more than 15 digits when varying the parameter $\lambda$ over a wide range.
\begin{table}[h]
\centering
\begin{tabular}{cc} \hline 
$\lambda$ [GeV] & $2\mathcal{\Re}(\mathcal{T}_{Tree}^{+}\mathcal{T}_{Loop})$+soft contribution\\ \hline   \hline
$10^{-17}$ & $-1.6743892369492021873805611201763810E^{-3}$\\ \hline
$10^{-19}$ & $-1.6743892369492020397654354220438766E^{-3} $\\ \hline
$10^{-21}$ & $-1.6743892369492020382892402083349623E^{-3} $\\ \hline
% $10^{-23}$ & $-1.6743892369492020382744348901161470E^{-3} $\\ \hline
\end{tabular}
\caption{\label{lambda} Test of the IR finiteness of the amplitude. 
In this table we take the non-linear gauge parameters to be $0$, 
$C_{UV}=0$ and the center-of-mass energy is $1$ TeV.}
\end{table}

The independence of the result on the five parameters  $\tilde{\alpha} , 
\tilde{\beta}, \tilde{\delta}, \tilde{\kappa}, \tilde\varepsilon$ is also 
checked. The result is presented in table~\ref{gauge}.
We find that the result is stable over more than 26 digits while varying the 
non-linear gauge parameters.
\begin{table}[h]
\centering
\small
\begin{tabular}{cc} \hline
$(\tilde{\alpha}, \tilde{\beta}, \tilde{\kappa},\tilde{\delta}, \tilde{\epsilon} ) $ 
& $2\mathcal{\Re}(\mathcal{T}_{Tree}^{+}\mathcal{T}_{Loop})$ \\ \hline  \hline 
$(0,0,0,0,0)$ &      $-5.3131630854021768119477116628317605E^{-3} $\\ \hline
$(1,2,3,4,5)$ &      $-5.3131630854021768119477116637537265E^{-3} $\\ \hline
$(10,20,30,40,50)$ & $-5.3131630854021768119477116582762373E^{-3} $\\ \hline
\end{tabular}
\caption{\label{gauge} Gauge invariance of the amplitude. 
In this table, we set $C_{UV} = 0$, the photon mass is 
$10^{-17}$GeV and a $1$ TeV center-of-mass energy.}
\end{table}

Finally, we check the stability of the result versus the soft photon cut 
parameter ($k_c$). This test includes both the soft photon and the hard 
photon contributions. The hard photon bremsstrahlung part is the process 
$e^+e^- \rightarrow t \bar{t} \gamma \gamma$. It is important to note that we have two photons
at the final state. One of them has to be applied an energy cut of $E_{\gamma}^{cut} \geq 10$ GeV 
and an angle cut of $10^{\circ}\leq \theta_{\gamma}^{cut} \leq 170^{\circ}$.
Another one is a hard photon with energy is greater than $k_c$ and smaller than the first photon's energy.
This part will be generated by the tree level version of 
GRACE~\cite{grace} with the phase space integration by BASES. The result is 
tested by changing the value of $k_c$ from $10^{-5}$ GeV to $0.1$ GeV. In 
table~\ref{kc} we find that the results are in agreement with an accuracy 
which is better than $0.1\%$ when we vary $k_c$.

\begin{table}[h]
\centering
\small
\begin{tabular}{cccc} \hline 
 $k_c $[GeV] & $ \sigma_{H} $  & $ \sigma_{S} $ &  $ \sigma_{S+H}$ \\ \hline  \hline
$10^{-5} $ & $ 4.172723E^{-02}  $  & $5.885469E^{-02}    $ &  $0.10058192$ \\ \hline  
$10^{-3} $ & $2.926684E^{-02} $  & $7.131737E^{-02} $ &  $0.10058421$ \\ \hline    
$10^{-1} $ & $1.678994E^{-02} $  & $8.377319E^{-02} $ &  $0.10056313$ \\ \hline   
\end{tabular}
\caption{\label{kc} Test of the $k_c$-stability of the result. We 
choose the photon mass to be $10^{-17}$ GeV and the center-of-mass energy 
is $1$ TeV. The second column presents the hard photon cross-section and 
the third column presents the soft photon cross-section. The final column 
is the sum of both.}
\end{table}

% The results have been checked carefully with three kinds of consistency test. 
We found that the numerical results are in good agreement when varying
$C_{UV}$, the gauge parameters, photon mass, and $k_c$. 
Hereafter, we set $\lambda=10^{-17}$ GeV, $C_{UV}=0$ 
and $\tilde{\alpha}= \tilde{\beta}= \tilde{\delta}= \tilde{\kappa}= 
\tilde\varepsilon=0$.
%--#] ee->ttg :
%--#[ Results :

\section{Results}

Our input parameters for the calculation are as follows. The fine structure 
constant in the Thomson limit is $ \alpha^{-1}=137.0359895$. The mass of 
the Z boson is $M_Z=91.187$ GeV. In the on-shell renormalization scheme 
we take the mass of the $W$ boson ($M_W$) as an input parameter. 
It will be derived through the electroweak 
radiative corrections to the muon decay width ($\Delta r$)~\cite{Hioki:1995ex} 
with $G_{\mu}=1.16639 \times 10^{-5}$ GeV$^{-2}$. 
Therefore, $M_W$ is a function of $M_H$. In this calculation, we take 
$M_H=120$ GeV and the numerical value of $M_W$ is $80.3759$ GeV. For the 
lepton masses we take $m_e=0.51099891$ MeV, $m_{\tau}=1776.82$ MeV and 
$m_{\mu}=105.658367$ MeV. For the quark masses we take $m_u=1.7$ MeV, 
$m_d=4.1$ MeV, $m_c=1.27$ GeV, $m_s=101$ MeV, $m_t=172.0$ GeV and 
$m_b=4.19$ GeV. We apply an energy cut of $E_{\gamma}^{cut} \geq 10$ GeV 
and an angle cut of $10^{\circ}\leq \theta_{\gamma}^{cut} \leq 170^{\circ}$ on the 
photon.

All numerical results are generated by the GRACE-Loop system. For 
$t\bar{t}$ production the results were first checked with the results in 
refs~\cite{Fujimoto:1987hu}, \cite{Fleischer:2002rn}, \cite{Fleischer:2003kk}. 
Then we use the values of the parameters above to produce the results of 
$t\bar{t}$ production in this paper and compare them with $t \bar{t} 
\gamma$ production.

\begin{figure}[h]
\begin{center}$
\begin{array}{cc}
\vspace*{-0.5cm}
e^- e^+ \rightarrow t \bar{t} &
e^- e^+ \rightarrow t \bar{t}  \gamma  \\
\includegraphics[width=3in,height=7.5cm,angle=-90]{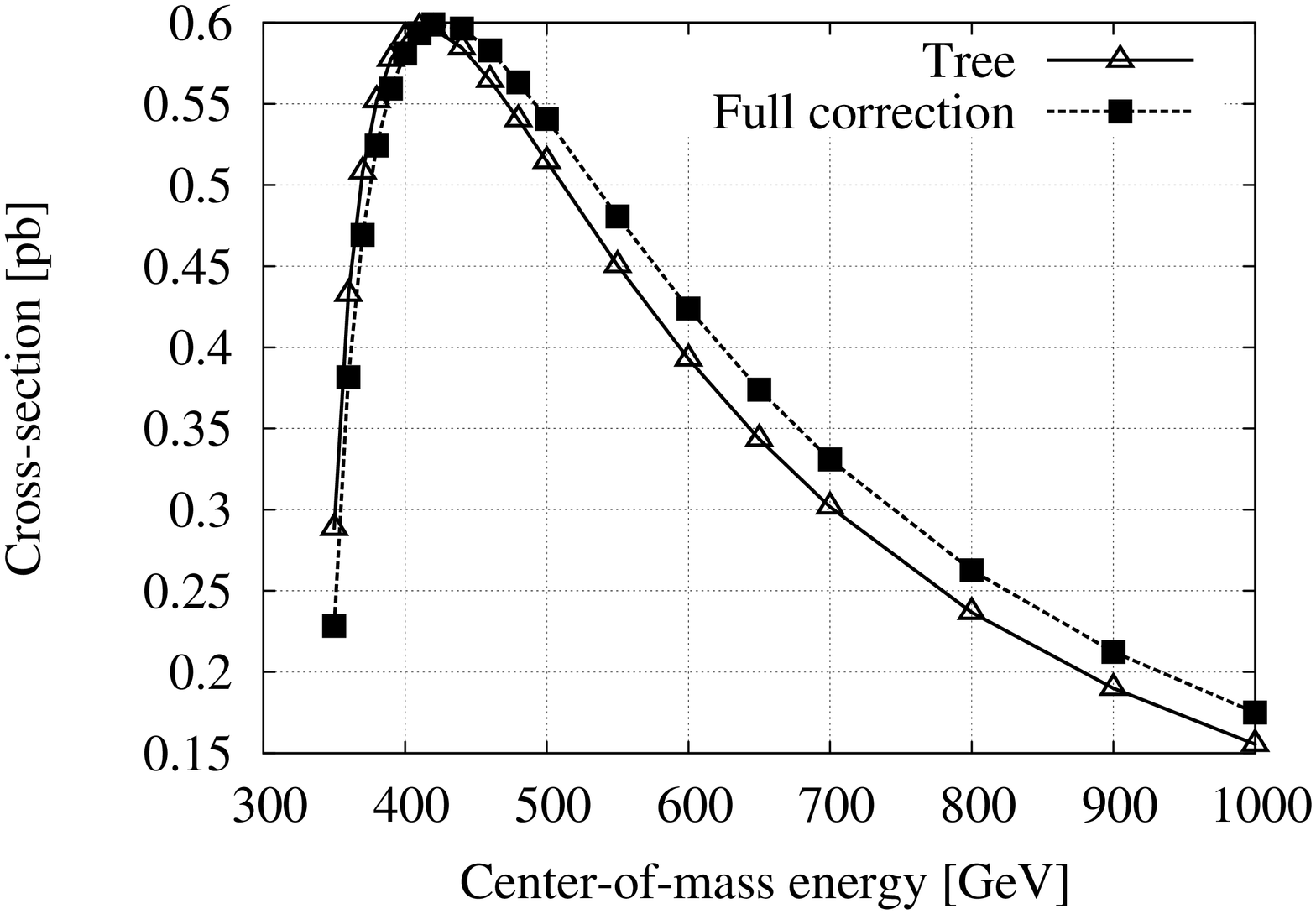} &
\includegraphics[width=3in,height=7.5cm,angle=-90]{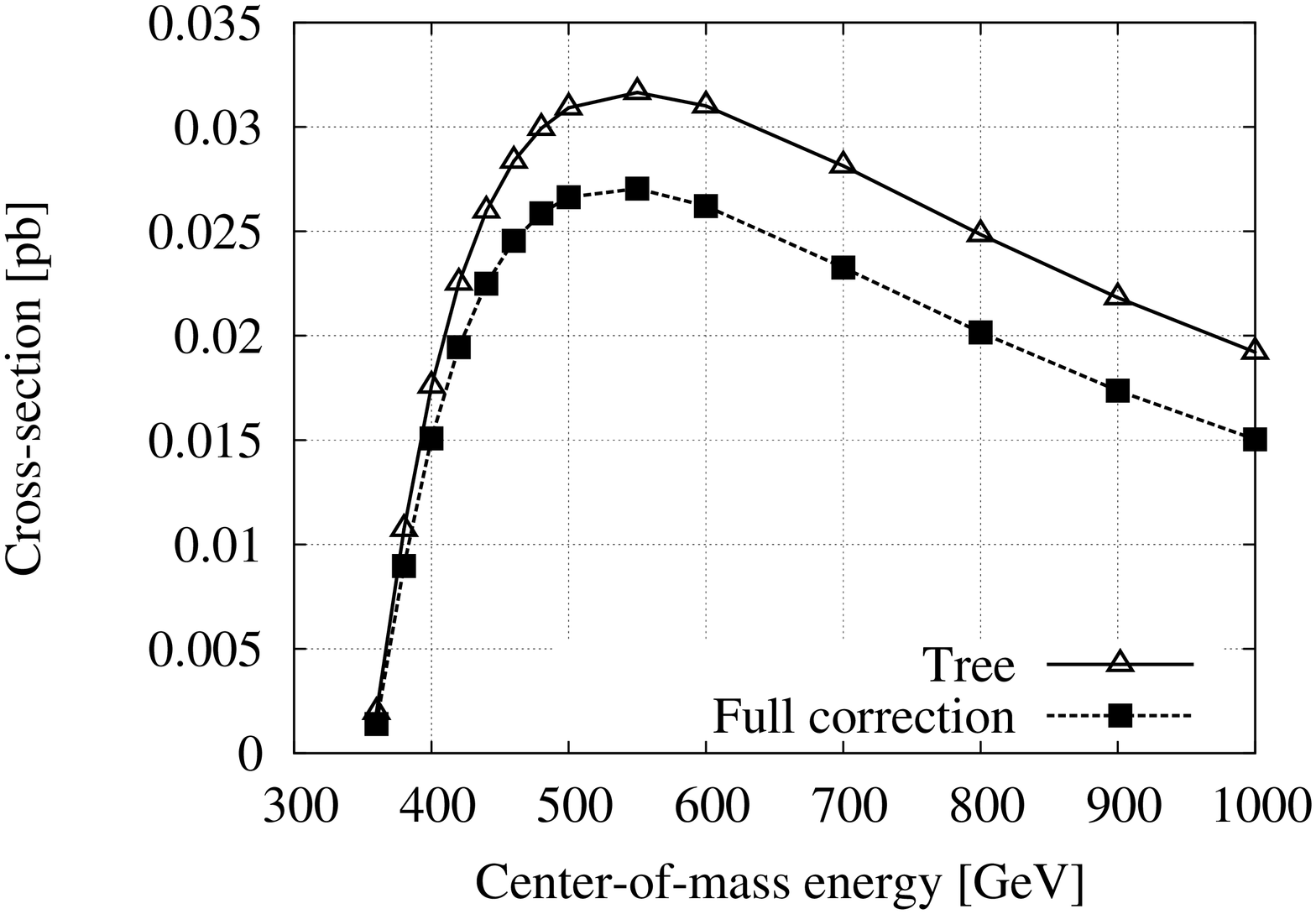}
\end{array}$
\end{center}
\caption{\label{cross-section} The total cross-section as a 
function of center-of-mass energy. The left figure is the result of $t 
\bar{t}$ production and the right figure shows the result of the $t \bar{t} 
\gamma$ reaction. The triangle points are the result of the tree 
level calculation while the rectangular points are the sum of the 
tree level calculation combined with the full one-loop electroweak 
radiative corrections. Lines are only guide for the eyes.}
\end{figure}

In Fig~\ref{cross-section} the total cross-section is a function of the 
center-of-mass energy $\sqrt{s}$. We vary the value of $\sqrt{s}$ from 
$360$ GeV to $1$ TeV. We find that the cross-section is largest near the 
threshold, $\sqrt{s}$ around $550$ GeV for $t \bar{t} \gamma$ production 
and $410$ GeV for $t \bar{t}$ production. The total cross-section of $t 
\bar{t} \gamma$ production is considerably less than $10\%$ of the total 
cross-section for the $t \bar{t}$ reaction. In addition we find a 
negative correction for $t \bar{t} \gamma$ production in 
contrast to the positive correction for $t \bar{t}$ production.

The full $\mathcal{O}(\alpha)$ electroweak corrections take into account 
the tree graphs and the full one-loop virtual corrections as well as the 
soft and hard bremsstrahlung contributions. The relative correction is 
defined as
\begin{eqnarray}
 \delta_{EW}=\frac{\sigma(\alpha)}{\sigma_{Tree} }-1.
\end{eqnarray}

In order to extract the genuine weak correction in the $G_{\mu}$ scheme, we 
first evaluate the QED initial radiative correction ($\delta_{QED}$). Applying the 
structure function method described in ref~\cite{supplement100}, 
$\delta_{QED}$ is defined as
\begin{eqnarray}
 \delta_{QED}&=&\frac{\sigma_{QED} -\sigma_{Tree} }{\sigma_{Tree}}, 
\end{eqnarray}
with
\begin{eqnarray} 
\sigma_{QED}(s)&=&\int\limits_{0}^1 dx \;\mathcal{H}(x,s)\; \sigma_{0}(s(1-x)), 
\end{eqnarray}
here $\mathcal{H}(x,s)$ is a radiator which is defined by formula (11.213) 
in ref~\cite{supplement100}:
\begin{eqnarray}
 \mathcal{H}(x,s) &=& \Delta \beta x^{\beta-1}-\beta (1-\frac{x}{2}) \nonumber\\
                  &+& \frac{\beta^2}{8} \Big [ -4(2-x)\ln x -\frac{1+3(1-x)^2}{x} \ln(1-x)-6+x \Big]
\end{eqnarray}
with $\beta=\frac{2 \alpha}{\pi}\Big ( \ln(\frac{s}{m_e^2})-1\Big )$ and 
$\Delta=1+\frac{\alpha}{\pi}\Big(\frac{3}{2}\ln(\frac{s}{m_e^2})
+\frac{\pi^2}{3}-2 \Big) $.\\

After obtaining the QED correction, we define the genuine weak 
correction in the $\alpha$ scheme:
\begin{eqnarray}
 \delta_{W}=\delta_{EW}-\delta_{QED}. 
\end{eqnarray}
Having subtracted the genuine weak correction in the $\alpha$ scheme, one can express the 
correction in the $G_{\mu}$ scheme. Next we subtract the universal weak 
correction which is obtained from $\Delta r$. The genuine weak correction 
in the $G_{\mu}$ scheme is defined by
\begin{eqnarray}
\delta^{G_{\mu}}_{W}= \delta_{W} -n \Delta r,
\end{eqnarray}
with $\Delta r=2.55\%$ for $M_H=120$ GeV and $n=3 (2)$ for $t \bar{t} \gamma$ (for $t \bar{t}$) production respectively.

\begin{figure}[ht]
\begin{center}$
\begin{array}{cc}
\vspace*{-0.5cm}
e^- e^+ \rightarrow t \bar{t}  &
e^- e^+ \rightarrow t \bar{t} \gamma \\
\includegraphics[width=3in,height=7.5cm,angle=-90]{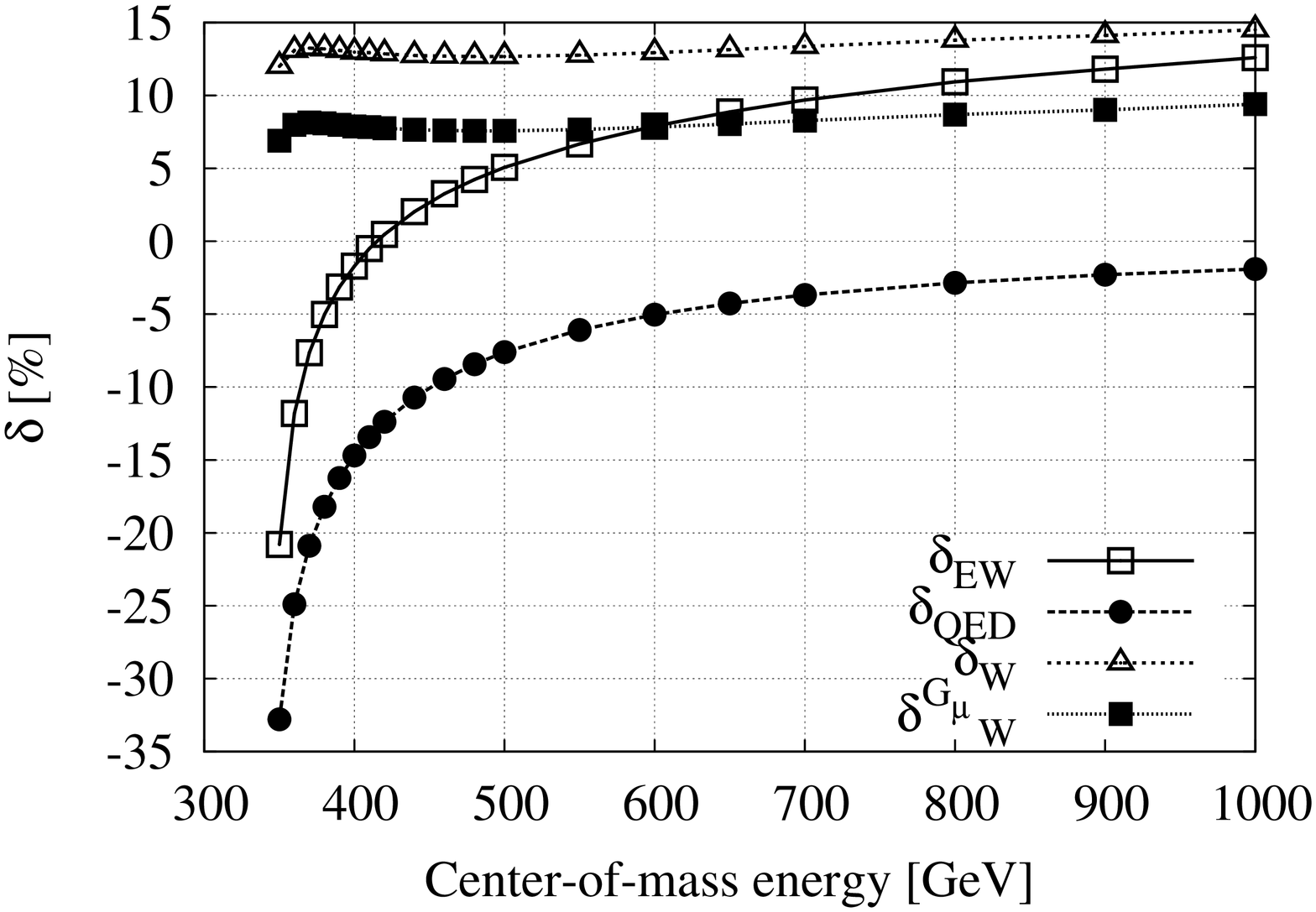} &
\includegraphics[width=3in,height=7.5cm,angle=-90]{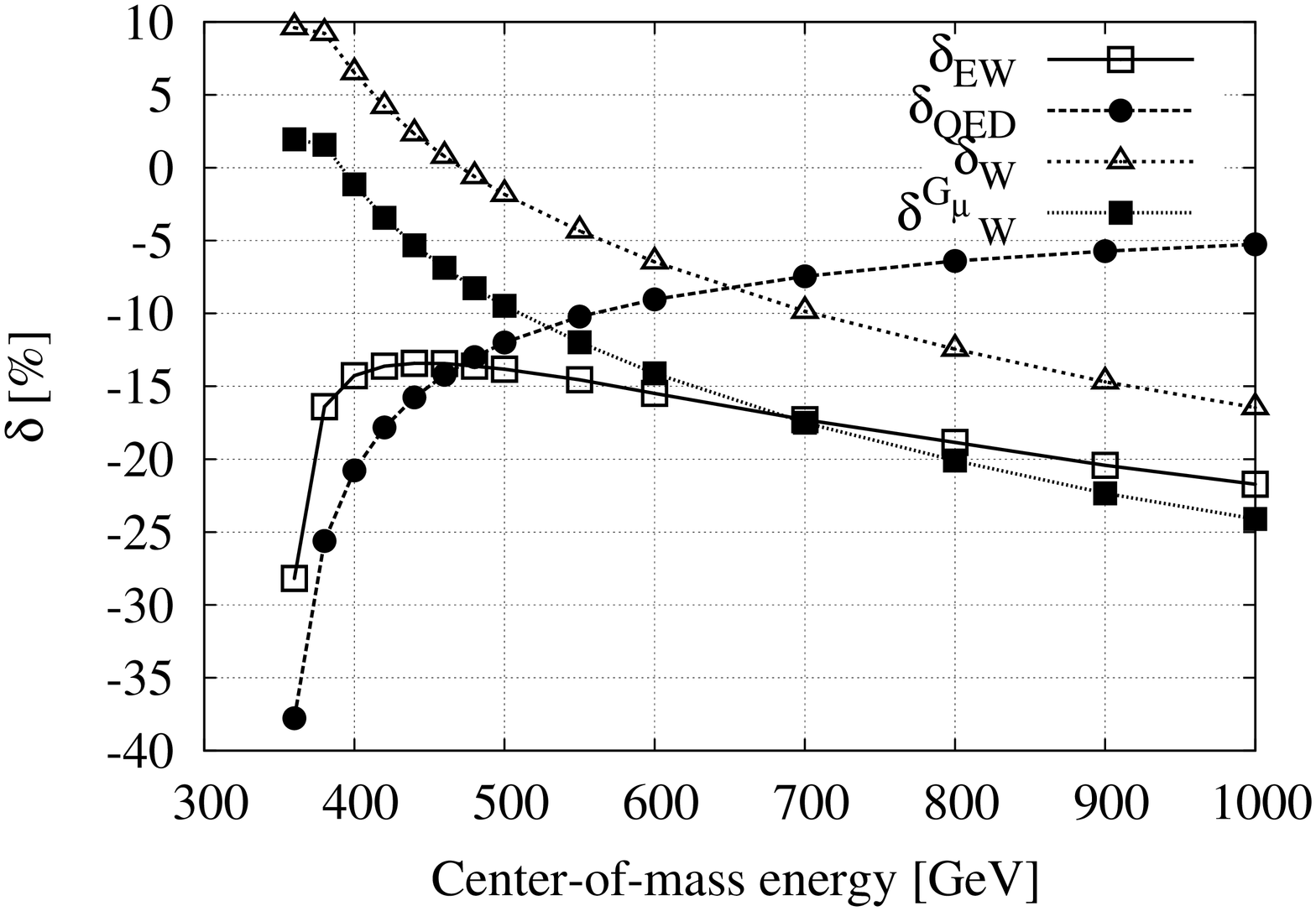}
\end{array}$
\end{center}
 \caption{\label{ew}The full electroweak correction and the 
genuine weak correction as a function of the center-of-mass energy. The left 
figure shows the results for $t \bar{t}$ production while the right figure 
shows the results for $t \bar{t} \gamma$ production. The circle 
points represent the QED correction, the empty rectangle 
points are the results for the full electroweak correction while the 
triangle points are the results for the genuine weak 
correction in the $\alpha$ scheme. The filled 
rectangle points represent the results of the genuine weak correction in 
the $G_{\mu}$ scheme. Lines are only guide for the eyes. }
\end{figure}

In Fig~\ref{ew}, we present the full electroweak correction and the 
genuine weak correction in both the $\alpha$ and the $G_{\mu}$ schemes for 
$t \bar{t}\gamma$ production as compared to $t \bar{t}$ production. These 
corrections are shown as a function of the center-of-mass energy, 
$\sqrt{s}$. We vary $\sqrt{s}$ from $360$ GeV to $1$ TeV. The figures show 
clearly that the QED correction is dominant in the low energy region. 
In the high energy region it is much smaller ($\sim 
-5\%$ at $1$ TeV). In contrast to the QED correction the weak correction in 
the $\alpha$ scheme is less than $10\%$ for low energies but reaches 
$-16\%$ at $1$ TeV center-of-mass energy. For $t \bar{t} \gamma$ 
production, we find that the value of the genuine weak correction in the 
$G_{\mu}$ scheme varies from $2\%$ to $-24\%$ over $\sqrt{s}$ from $360$ GeV to $1$ TeV.

Now we turn our attention to the forward-backward asymmetry $A_{FB}$. This 
quantity is defined as
\begin{eqnarray}
 A_{FB} & = & \frac{\sigma(0^{\circ}\leq \theta_{t}\leq 90^{\circ})
   -\sigma(90^{\circ}\leq \theta_{t}\leq 180^{\circ})}
   {\sigma(0^{\circ}\leq \theta_{t}\leq 90^{\circ})
   +\sigma(90^{\circ}\leq \theta_{t}\leq 180^{\circ})},
\end{eqnarray}
with $\theta_{t}$ the angle of the top quark.

\begin{figure}[ht]
\begin{center}$
\begin{array}{cc}
\vspace*{-0.5cm}
e^- e^+ \rightarrow t \bar{t} &
e^- e^+ \rightarrow t \bar{t} \gamma \\
\includegraphics[width=3in,height=7.5cm,angle=-90]{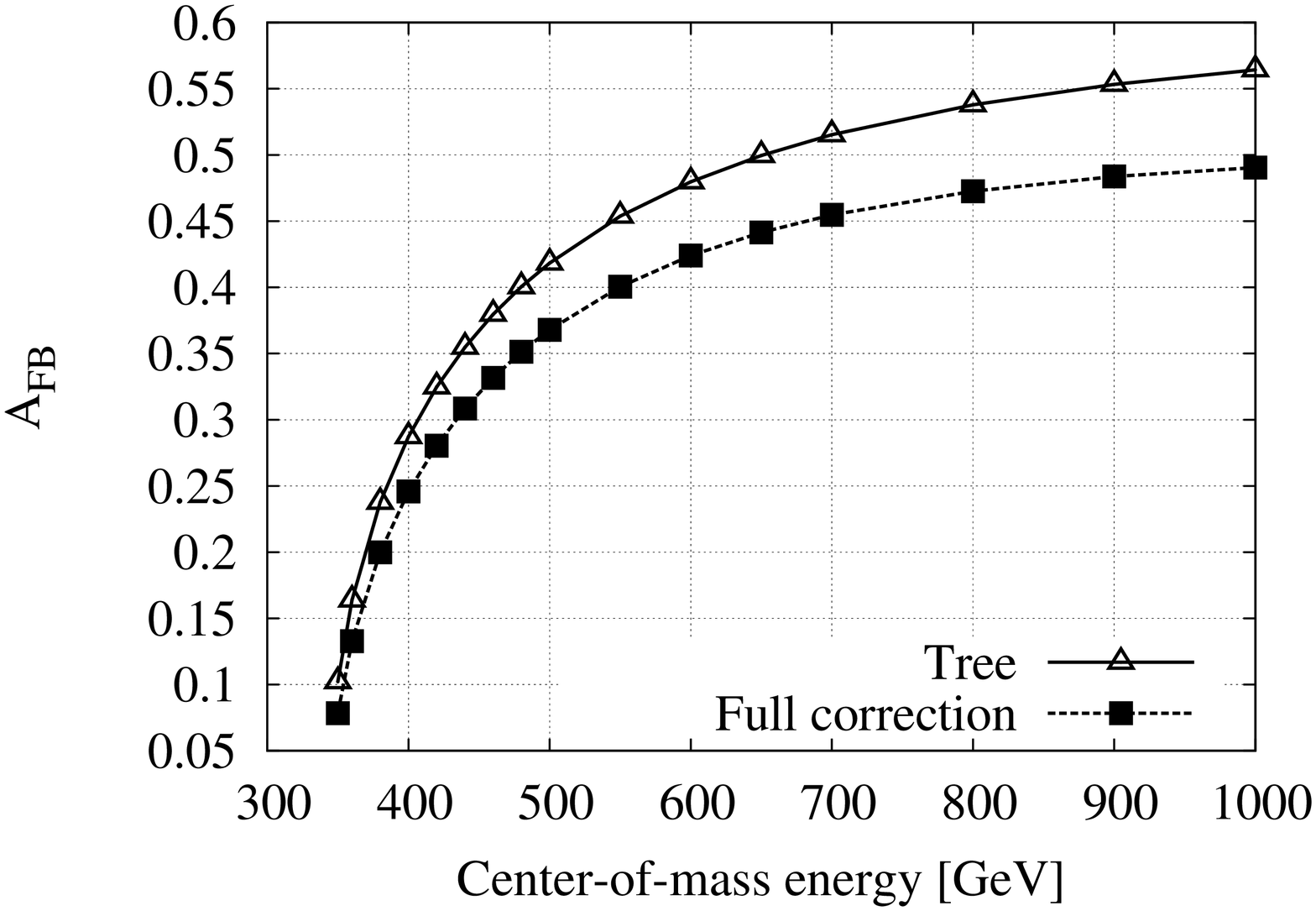} &
\includegraphics[width=3in,height=7.5cm,angle=-90]{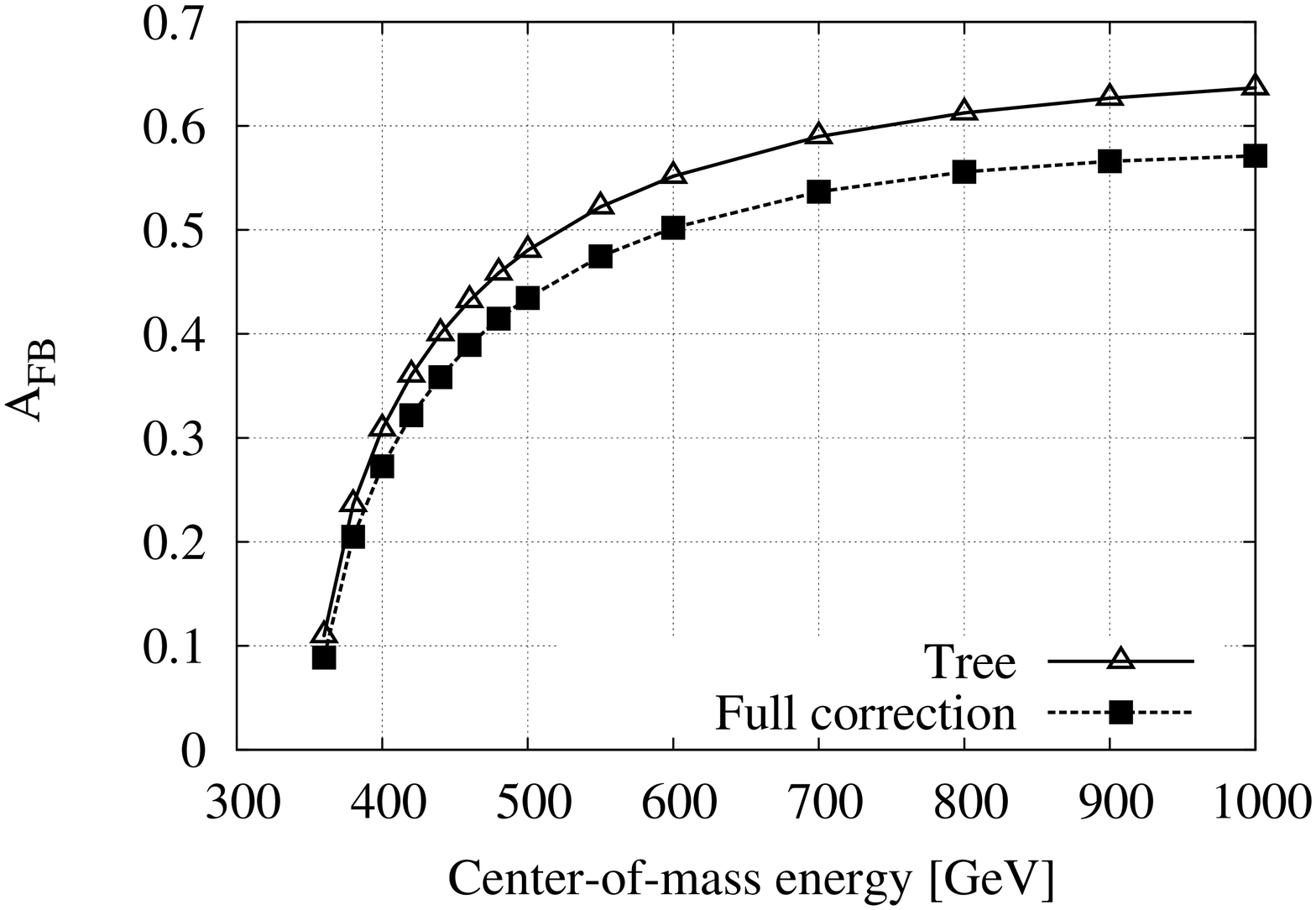}
\end{array}$
\end{center}
 \caption{\label{af} The top quark forward-backward asymmetry 
as a function of the center-of-mass energy. Left figure is the results for $t 
\bar{t}$ production and right one is the results for $t \bar{t} \gamma$ 
production. The triangle points 
represent the tree level results and the rectangle points are the 
results including the full radiative corrections. Lines are only guide for the eyes.}
\end{figure}

Fig~\ref{af} shows the results for $A_{FB}$ as a function of the 
center-of-mass energy. The figures show clearly that the top quark 
asymmetry in the full results is smaller than the asymmetry at the tree 
level results only.

\begin{figure}[ht]
\begin{center}$
\begin{array}{cc}
\includegraphics[width=3in,height=7.5cm,angle=-90]{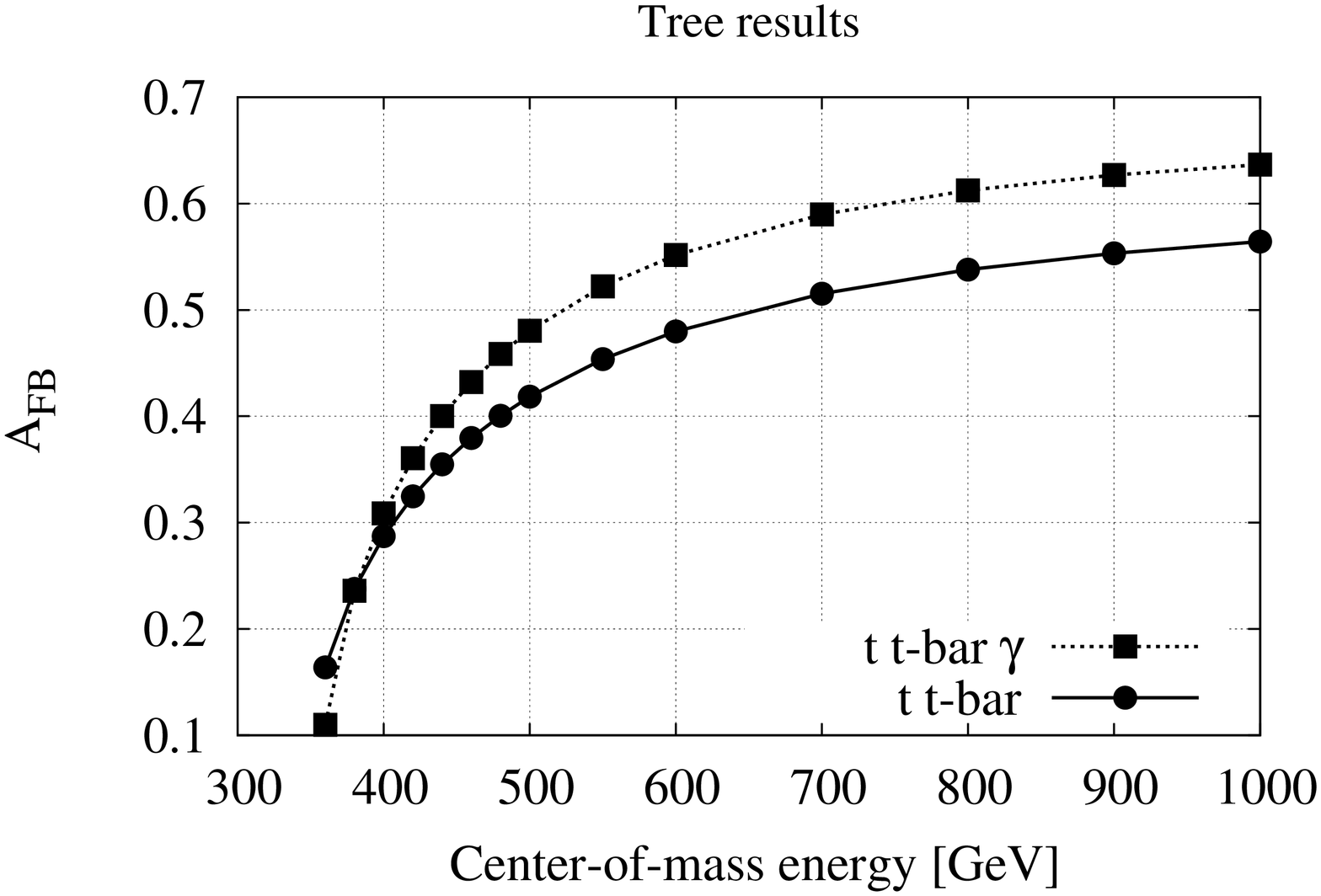} &
\includegraphics[width=3in,height=7.5cm,angle=-90]{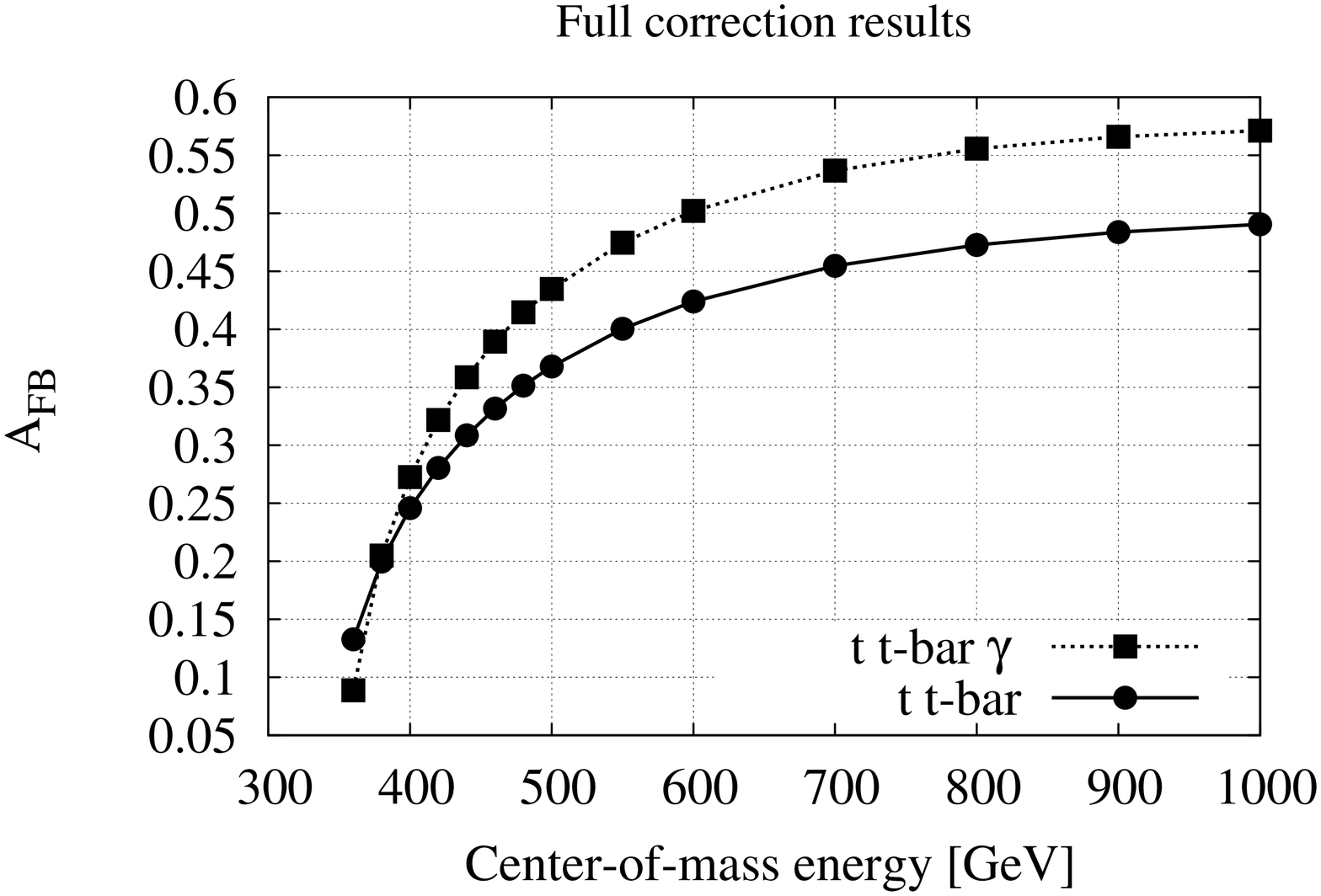}
\end{array}$
\end{center}
 \caption{\label{Afbcompare} The value of the top quark 
asymmetry in $t\bar{t}\gamma$ production as compared to $t\bar{t}$ 
production. The rectangular points (circle points) represent  the 
result for $t\bar{t}\gamma$ production ($t\bar{t}$ production) 
respectively. Lines are only guide for the eyes.}
\end{figure}

In Fig~\ref{Afbcompare} we compare the values of $A_{FB}$ in $t \bar{t} 
\gamma$ production directly with its value for $t \bar{t}$ production. From 
the figures, we find that $A_{FB}$ in $t \bar{t} \gamma$ production is 
larger than $A_{FB}$ in $t \bar{t}$ production. This is the most important 
result of the paper. The effect should be clearly observable at the ILC.
%
%	The original remark about testing this at the Tevatron is probably out 
%	of context because there the signal is due to different reactions.
%

%  \clearpage

%--#] Results :
%--#[ Conclusions :

\section{Conclusions} 

We have presented the full $\mathcal{O}(\alpha)$ electroweak 
radiative corrections to the process $e^+e^- \rightarrow t \bar{t} \gamma$ and
$e^+e^- \rightarrow t \bar{t}$
at ILC. The calculations were done with the help of 
the GRACE-Loop system.

GRACE-Loop have implemented a generalised non-linear gauge fixing condition which includes five
gauge parameters. With the UV, IR finiteness and gauge parameters independence checks, the system
provide a powerful tool to test the results in the consistency way. In the numerical checks 
of this calculation, we find that the results are numerically stable when quadruple precision is used.

We find that the numerical value of the genuine weak corrections in $G_{\mu}$ scheme varies 
from $2\%$ to $-24\%$ in the range of center-of-mass energy from $360$ GeV to $1$TeV.
 We also obtain a large value for the top quark forward-backward 
asymmetry in the $t \bar{t} \gamma$ process as compared with the one in $t 
\bar{t}$ production.

We also introduce the axial gauge for 
the external photon in the GRACE-Loop system. It helps to avoid a large 
numerical cancellation problem. 
This is very useful when calculating Bhabha scattering at small angle and 
energy cuts of the final state particles. Bhabha scattering and related 
processes are not only used as luminosity monitor, but also play an 
important role as backgrounds for the process $e^- e^ + \rightarrow 
\tilde{\chi}^- \tilde{\chi}^+ \gamma$, which is a very interesting reaction 
for the search for dark matter. We will address it in a future publication.
% This is however not the topic of the 
% current paper. 

In addition, this calculation will provide a framework for calculating the 
full $\mathcal{O}(\alpha)$ electroweak radiative corrections for the 
process $e^+e^- \rightarrow e^+e^- \gamma$. This reaction and these 
corrections will play an important role at future $e^+e^-$ colliders like 
the ILC.

\section*{ Acknowledgments} We wish to thank Dr. F.~Yuasa and Dr. N.~Watanabe for 
valuable discussions and comments. This work was supported by JSPS KAKENHI 
Grant Number 20340063. The work of T.U. was supported by the DFG through SFB/TR 9 ``Computational 
Particle Physics''.

%--#] Conclusions :
%--#[ Bibliography :

%--#] Bibliography :

\end{document}